\documentclass[preprint,12pt]{elsarticle}

\usepackage{lineno,hyperref}
\modulolinenumbers[1]

\journal{N.I.M. A}\pdfoutput=1
\usepackage{graphicx}
\usepackage{color}
\usepackage{mathtools}
\usepackage{subfigure}
\usepackage{hyperref}

\usepackage{natbib}

\begin{document} 
\begin{frontmatter}

\title{Simultaneous scintillation light and charge readout of a pure argon filled Spherical Proportional Counter}

\author[cenbg]{R.~Bouet}
\author[cppm]{J.~Busto}
\author[cenbg,subatech]{V.~Cecchini \corref{mycorrespondingauthor}}
\cortext[mycorrespondingauthor]{Corresponding author}
\author[cenbg]{C.~Cerna}
\author[lsm]{A.~Dastgheibi-Fard}
\author[cenbg]{F.~Druillole}
\author[cenbg]{C.~Jollet}
\author[cenbg]{P.~Hellmuth}
\author[bir]{I.~Katsioulas}
\author[bir,cea]{P.~Knights}
\author[cea]{I.~Giomataris}
\author[cea]{M.~Gros}
\author[subatech]{P.~Lautridou}
\author[cenbg]{A.~Meregaglia}
\author[cea] {X.~F.~Navick}
\author[bir]{T.~Neep}
\author[bir]{K.~Nikolopoulos}
\author[cenbg]{F.~Perrot}
\author[cenbg]{F.~Piquemal}
\author[cenbg]{M.~Roche}
\author[cenbg]{B.~Thomas}
\author[bir]{R.~Ward}
\author[lsm]{M.~Zampaolo}
\address[cenbg]{CENBG, Universit\'{e} de Bordeaux, CNRS/IN2P3, 33175 Gradignan, France}
\address[cppm]{CPPM, Universit\'{e} d'Aix-Marseille, CNRS/IN2P3, F-13288 Marseille, France}
\address[subatech]{SUBATECH, IMT-Atlantique, Universit\'{e} de Nantes, CNRS/IN2P3, France}
\address[lsm]{LSM, CNRS/IN2P3, Universit\'{e} Grenoble-Alpes, Modane, France}
\address[bir]{School of Physics and Astronomy, University of Birmingham, B15 2TT, UK}
\address[cea]{IRFU, CEA, Universit\'{e} Paris-Saclay, F-91191 Gif-sur-Yvette, France}

\begin{abstract}
The possible use of a Spherical Proportional Counter for the search of neutrinoless double beta decay is investigated in the R2D2 R\&D project. Dual charge and scintillation light readout may improve the detector performance.
Tests were carried out with pure argon at 1.1 bar using a 6$\times$6~mm$^2$ silicon photomultiplier. Scintillation light was used for the first time to trigger in a spherical proportional counter. The measured drift time is in excellent agreement with the expectations from simulations.
Furthermore the light signal emitted during the avalanche development exhibits features that could be exploited for event characterisation.
\end{abstract}


\end{frontmatter}

\section{Introduction}
The use of Spherical Proportional Counters (SPC) in direct dark matter searches has been going on for a decade within the NEWS-G collaboration~\cite{Giomataris:2008ap,Gerbier:2013vta,Gerbier:2014jwa,Arnaud:2018bpc}. More recently, the possibility to use such a technology to search for $\beta\beta0\nu$ decay has been investigated~\cite{Meregaglia:2017nhx} and an R\&D programme is ongoing~\cite{Bouet:2020lbp}.
The aim is to tackle the various technical challenges to be faced for the construction of a large SPC, up to 1~m radius, filled with $^{136}$Xe at 40 bar and simultaneously to demonstrate that energy resolution of 1$\%$ FWHM, and the recognition and radial localization of the two searched-for beta tracks are possible with this very simple setup. The proposed R\&D programme addresses the different topics sequentially, before designing the final apparatus optimized for the targeted physics program.

So far, very encouraging results on the energy resolution (1.1$\%$ FWHM) have been obtained in Ar at 1 bar, and a coarse localization of the interactions within the detector seemed possible~\cite{Bouet:2020lbp}. In parallel, the NEWS-G collaboration has demonstrated with SPC low-pressure operations that identification and analysis of events initiated by 2 distinct electrons was routinely reached~\cite{MPTAUP,NKTAUP}.
Likewise, the search for KK axions in their expected di-photon decay channel was carried out~\cite{FV}. These experimental advances suggests that the identification of double traces induced by a double beta decay would be possible.

In this study, our objective is to verify the possibility of using a pure noble gas and to characterise the detector performance obtained at atmospheric pressure before turning to higher pressure. The light emission being one of the distinguishing characteristics of the excitation in a noble gas, we implemented a minimal system to detect scintillation light, and coincidences with the SPC detector signal were observed.

In this paper the experimental setup is described with emphasis on the light readout system. We compare between observed data and simulation and discuss the light emitted in the avalanche process. We also evaluate the impact of quencher-free gas on the charge signal shape and resolution.

\section{Experimental setup}


\begin{figure}[t]
\centering
\includegraphics[width=0.55\textwidth]{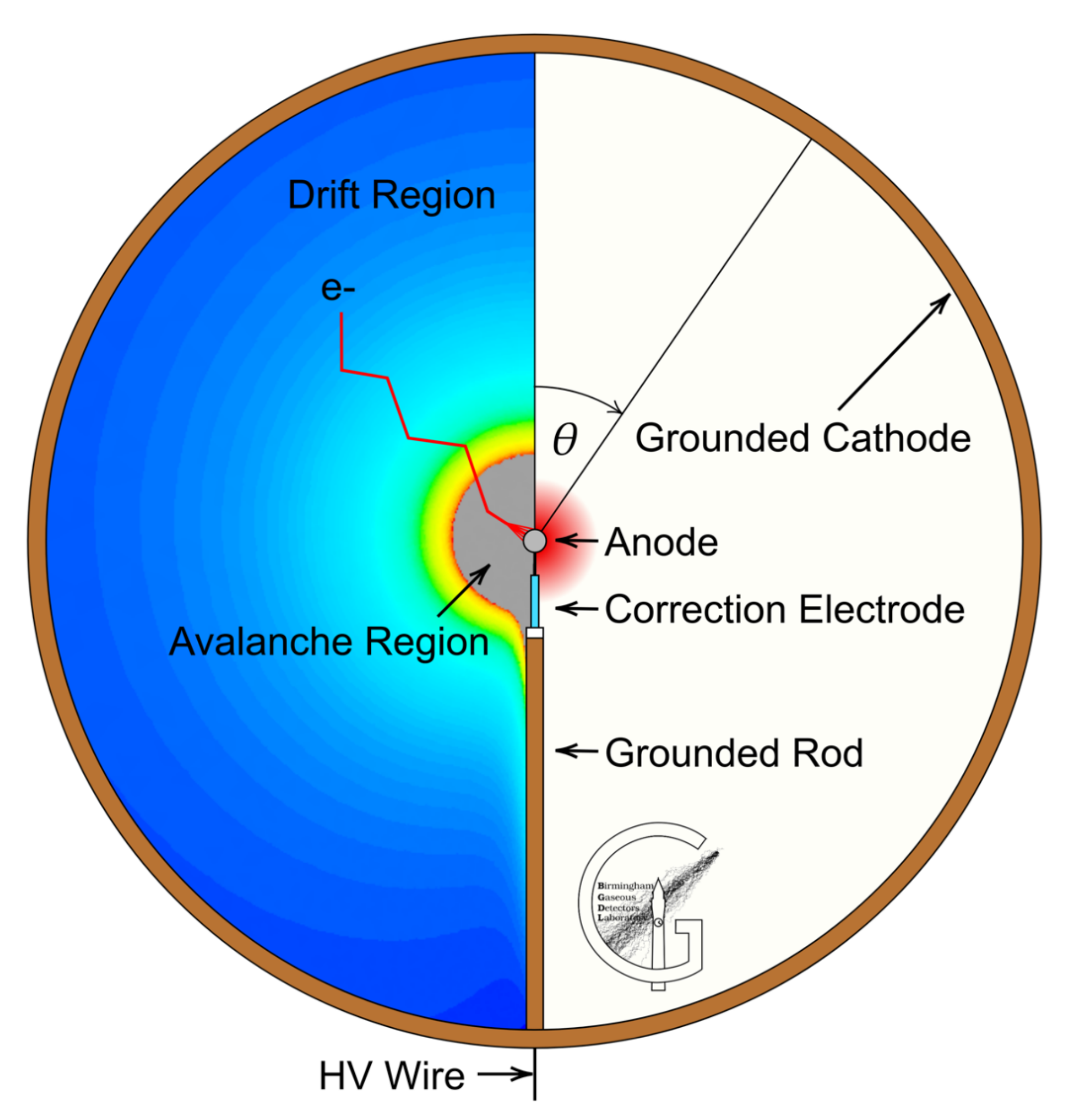} 
\caption{{\it Operating principle of the SPC detector taken from Ref.~\cite{Katsioulas:2020ycw}.}\label{fig:fig1}}
\end{figure}

The detector operating principle consists of a spherical grounded vessel, filled with gas and with a central anode at a positive high voltage as shown in Fig.~\ref{fig:fig1}. Particles passing through the gas ionize it and produce electrons which drift under the influence of the electric field to the central anode. When the electrons reach a distance of few mm from the central anode, depending on the high voltage (HV), they create an avalanche leading to the positive ions signal. More details can be found in Refs.~\cite{Giomataris:2008ap,Savvidis:2016wei}.

The prototype built at CENBG, in the framework of the R2D2 R\&D, consists of a  spherical stainless steel vessel of 20~cm radius with a central anode of 2~mm diameter. At the bottom of the detector there is an opening which is used to insert a $^{210}$Po source emitting $\alpha$ particles with an energy of 5.3~MeV. This prototype was used to assess the energy resolution in Ar at different pressures and further details on the setup can be found in Ref.~\cite{Bouet:2020lbp}.

\subsection{Light readout system}

\begin{figure}[t]
\centering
\includegraphics[width=0.55\textwidth]{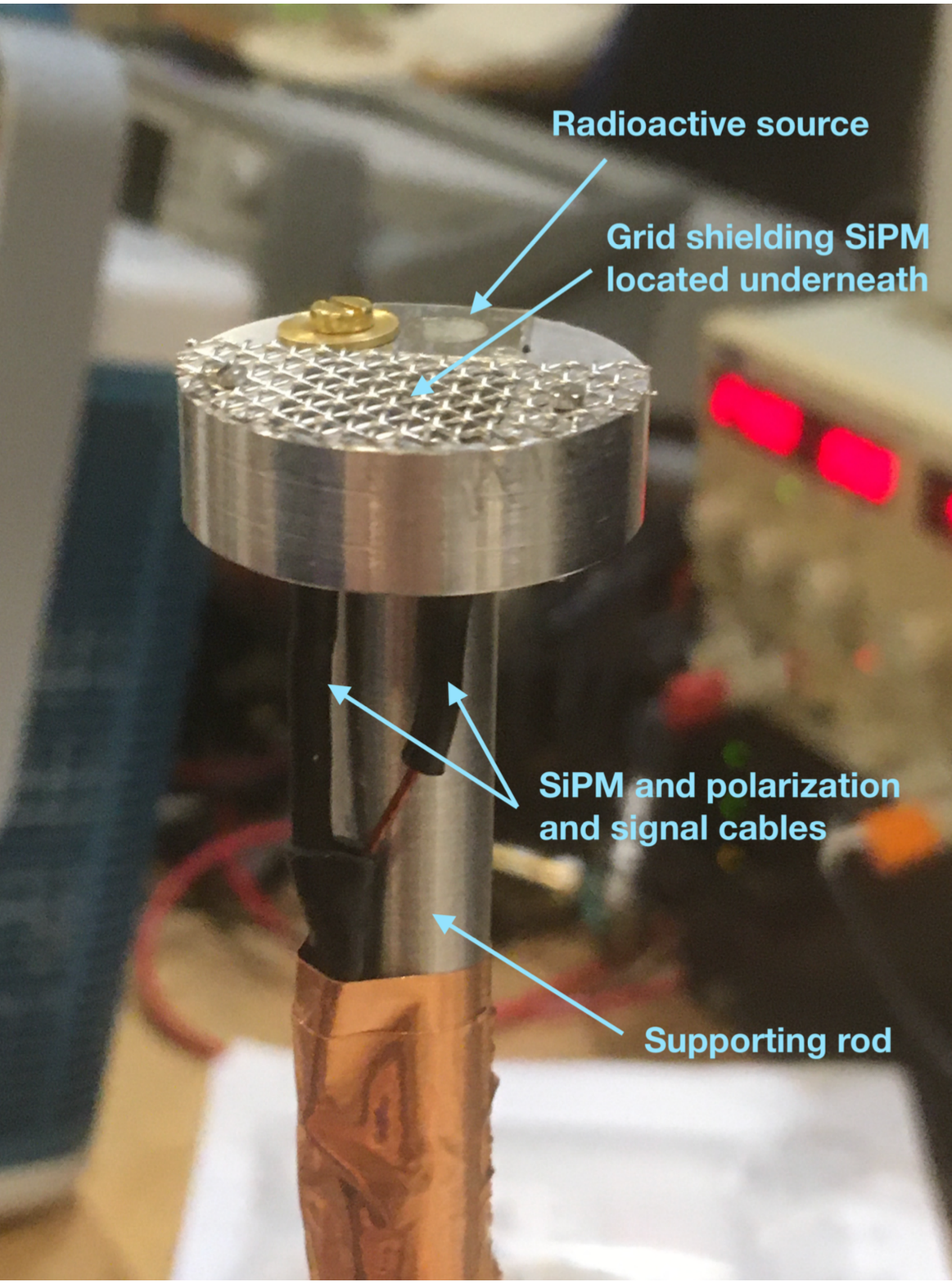} 
\caption{{\it SiPM setup.}\label{fig:fig2}}
\end{figure}

\begin{figure} [t]
\centering
\subfigure[\label{fig:fig32}]{\includegraphics[height=8.5
cm]{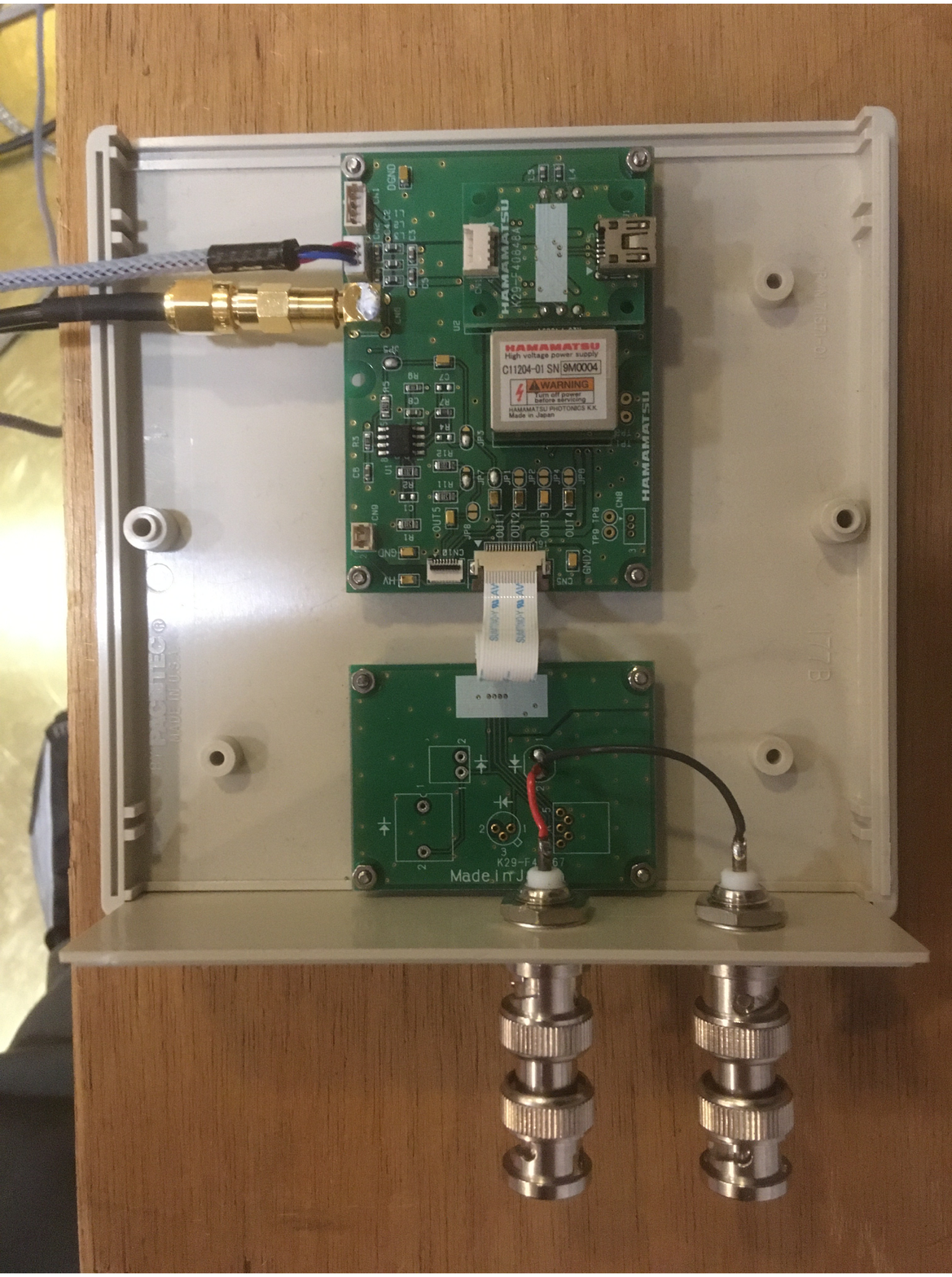}}
\subfigure[\label{fig:fig31}]{\includegraphics[height=8.5
cm]{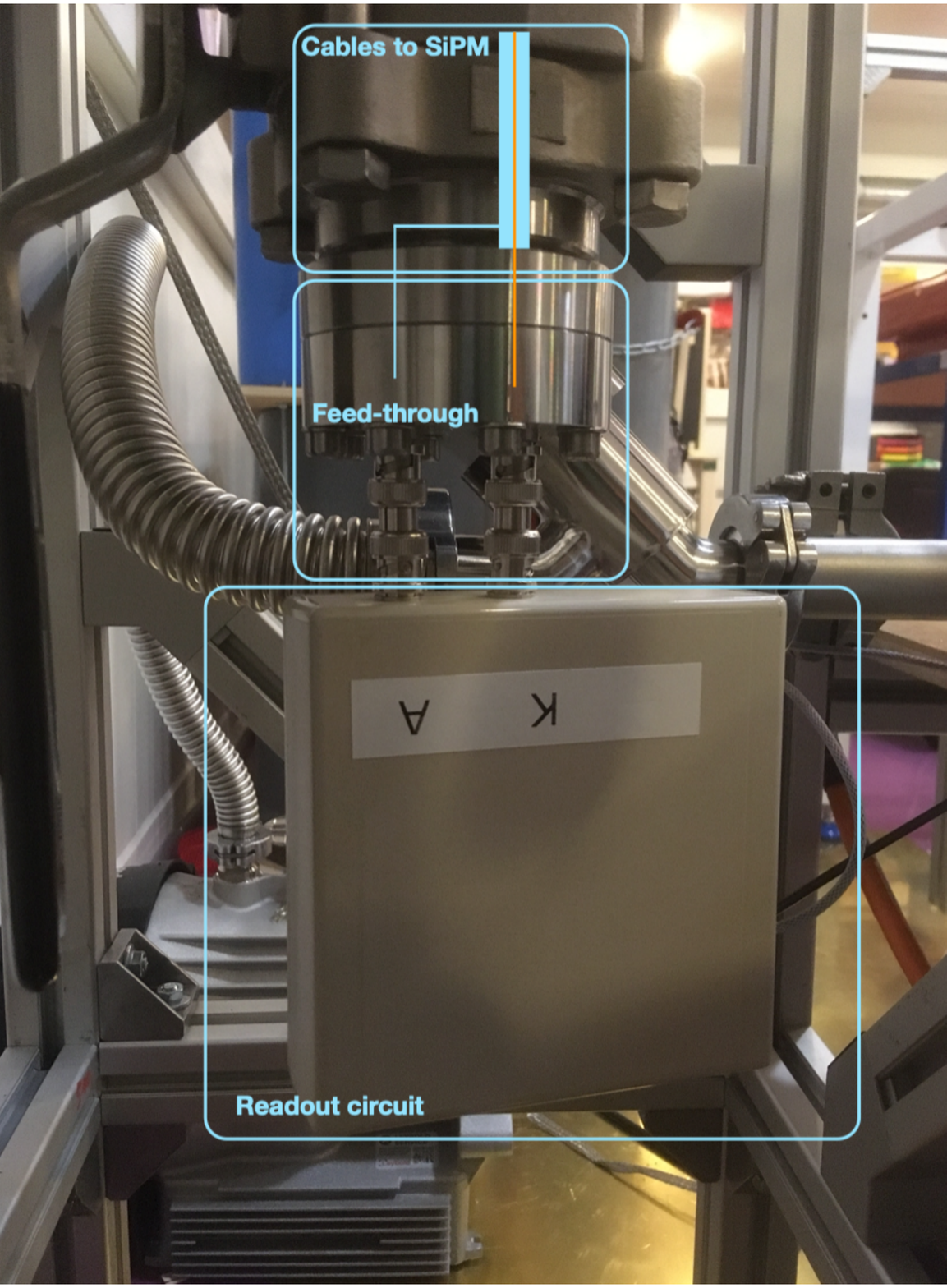}} \caption{{\it Picture of the open readout box with the MPPC C12332-01 readout circuit (\subref{fig:fig32}) and SiPM readout circuit setup (\subref{fig:fig31}).}}
\label{fig:fig3}
\end{figure}

\begin{figure} [t]
\centering
\subfigure[\label{fig:fig41}]{\includegraphics[height=4.5
cm]{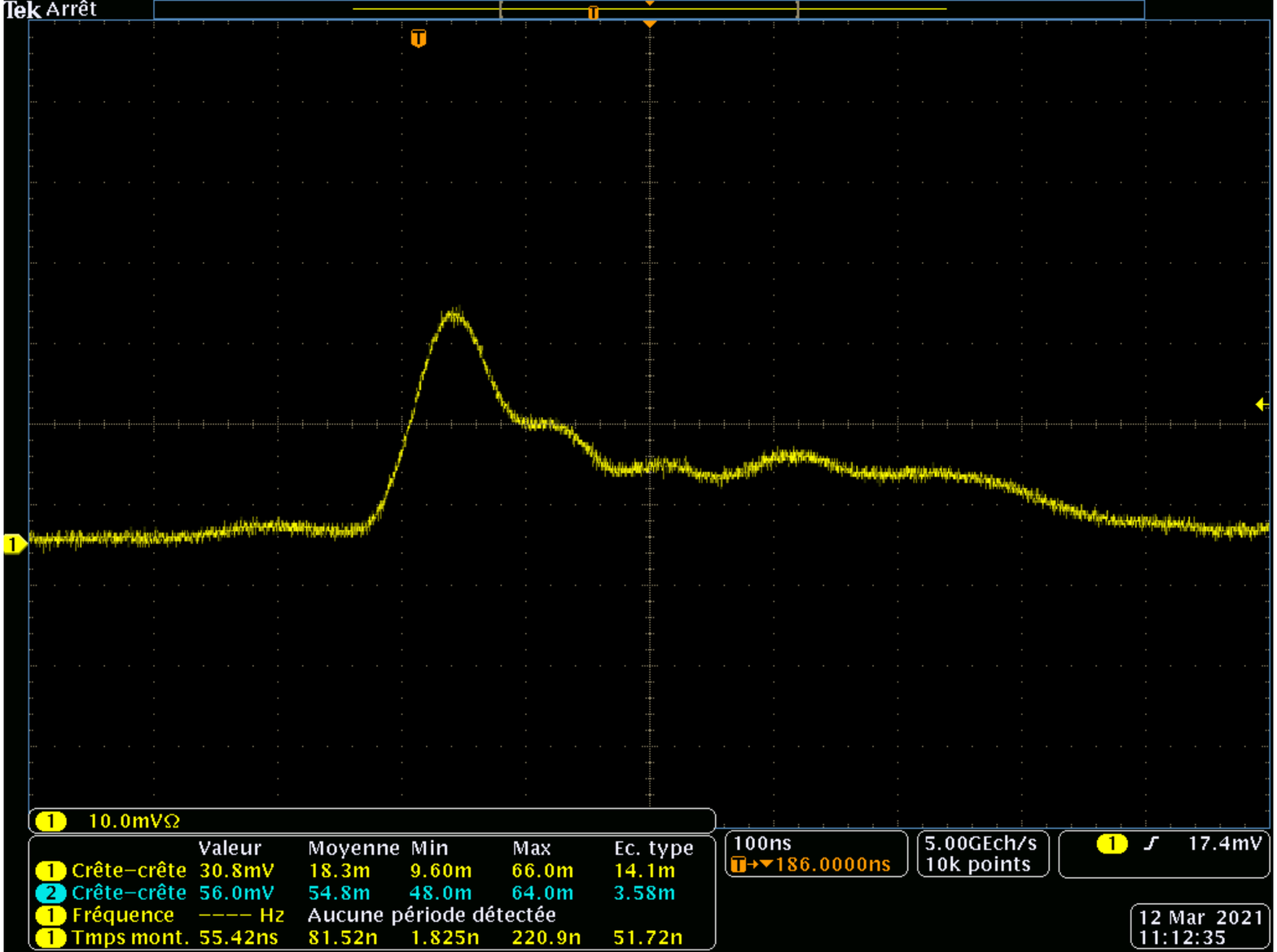}}
\subfigure[\label{fig:fig42}]{\includegraphics[height=4.5
cm]{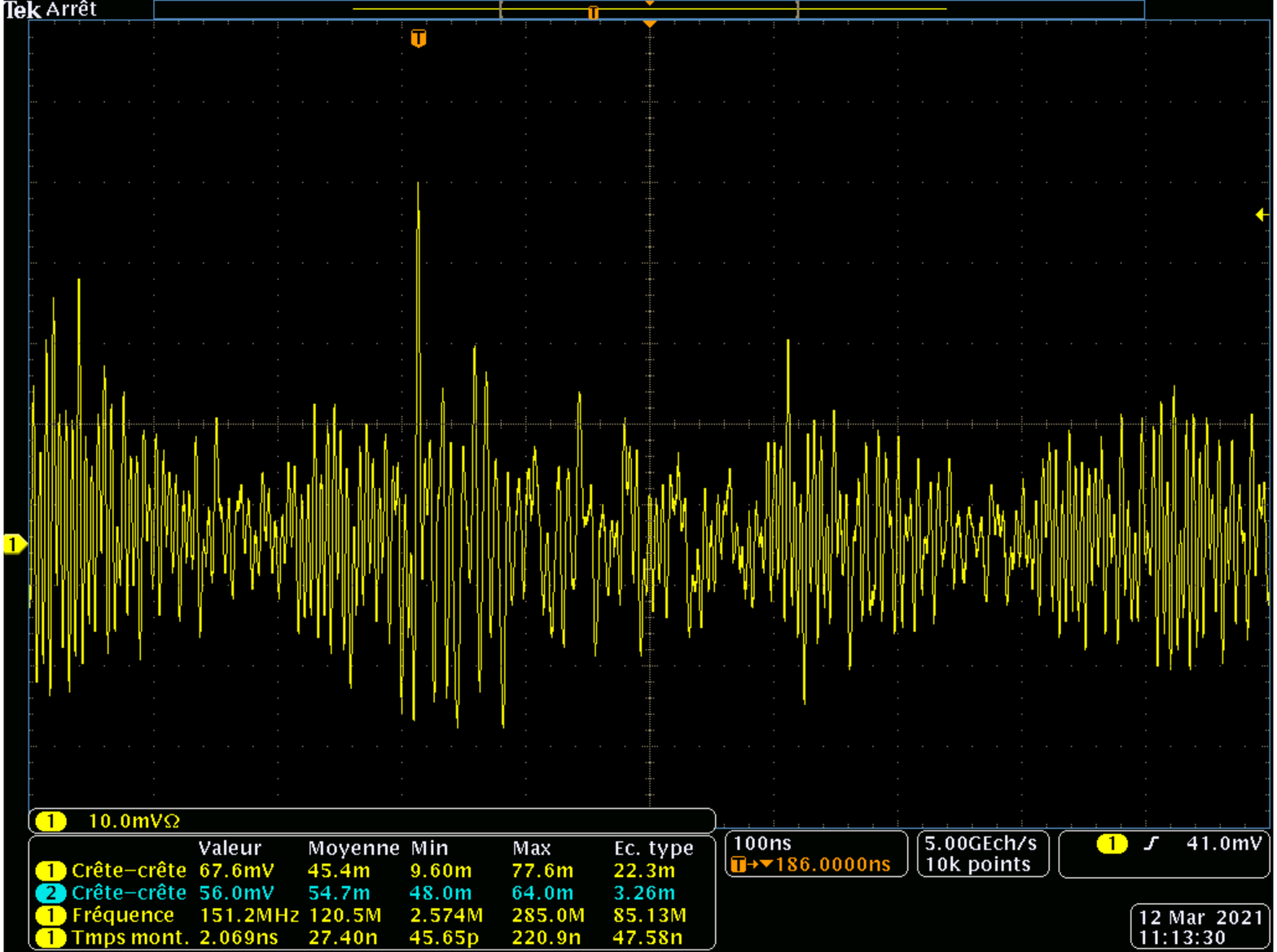}} \caption{{\it Signal of SiPM with a threshold of about 10 photoelectrons with (\subref{fig:fig41}) and without (\subref{fig:fig42}) the CLPFL-0010-BNC  CRYSTEK low-pass filter of 10~MHz.}}
\label{fig:fig4}
\end{figure}

The light yield of Ar gas is 14700 photons per MeV, assuming that electronegative impurities are kept at the ppm level~\cite{Amsler:2007gs}. In our case the detector was pumped to a vacuum of $5 \times 10^{-5}$~mbar, and the use of pure Ar (\textit{i.e.} purity at the level of 99.9999\% given by supplier) allowed to reach a purity sufficient for light detection and a homogeneous gain, \textit{i.e.} independent of $\alpha$ track radial distance. No further purification of the gas was attempted. Data taking was started 6.5 h after filling the SPC to ensure stable operation. The detector stability was monitored to ensure that the gain response remains constant over time.

To observe argon scintillation light a photodetector sensitive to 128~nm photons is required. Generally, liquid argon experiments aiming at detecting scintillation light, exploit Tetraphenyl-Butadiene (TPB) coating of surfaces to shift the wavelength of the light into a range visible to Photomultiplier Tubes (PMTs). Such a procedure is however delicate and our chosen solution was to use commercial VUV silicon photomultipliers (SiPM) from Hamamatsu. Indeed SiPM S13370, with a size of $6\times6$~mm$^2$ is a product of the VUV4 family with a photon detection efficiency (PDE) of 14\% at 128~nm.

In the R2D2 setup the SiPM was mounted on the source support as shown in Fig.~\ref{fig:fig2}. Such a position has several advantages including the maximisation of the number of detected photons emitted by the $\alpha$ track (tracks passing in front of the SiPM), and the ease of cabling power supply and signal readout through the feedthrough located at the bottom of the detector. There is, however, a drawback: the SiPM had to be operated at a voltage of 55~V and it is located in a region where the electric field was weak, at the level of 0.1 V/cm. Since the SiPM surface was at 55~V most of the electrons from the $\alpha$-particle track would have drifted towards the SiPM instead of drifting towards the central anode. To overcome this issue a grid with holes of 1~mm and an optical transparency of 65\%, was installed in front of the SiPM acting as a Faraday cage. Dedicated tests showed that turning the SiPM on and off had no impact on the signal and on the energy resolution proving the efficiency of the grid in shielding the SiPM field.

Since the SiPM is operated at room temperature a self-trigger rate at the level of kHz is expected for thresholds up to 4--5 photoelectrons (p.e.) depending on the SiPM. This, however, was not an issue for this application where the threshold could be raised to about 10 photoelectrons. A further challenge was related to the electronic noise due to the cables connecting the SiPM to the Hamamatsu readout circuit. Such a circuit (MPPC C12332-01), self-regulated with respect to temperature variations, was used to read out the SiPM signal, but it was located outside the detector, as shown in Fig.~\ref{fig:fig32}. This choice was  imposed by the requirement to minimize the presence of material inside the detector. However, to reduce the noise, the distance between the SiPM and the readout circuit was minimized: the circuit was connected directly at the feedthrough as shown in Fig.~\ref{fig:fig31}. Inside the detector, between the feedthroughs and the SiPM there was a distance of about 30~cm which could not be reduced. Several options were explored, but the solution put forward was to use a single coaxial cable, using the inner part of the cable for the signal transmission and the outer part for the power supply of the SiPM.

The SiPM signal is relatively fast and its width is of about 200~ns. A high frequency noise due to the distance between the SiPM and the readout circuit was observed and eliminated with a low-pass filter of 10~MHz (CLPFL-0010-BNC) from CRYSTEK (Fig.~\ref{fig:fig4}). Such a configuration allowed reaching the best experimental condition in order to look at the SiPM signal in coincidence with the SPC signal.

\section{Data analysis}
\label{sec:Analysis}

Different runs were taken at different pressures and HV values. For pressures below 500~mbar $\alpha$-particle tracks are too long (i.e. more than 6~cm) to give enough detectable light on the SiPM, and therefore the runs used in the presented analysis were all taken at the maximal allowed detector pressure of 1.1~bar. Two HV regimes were investigated: HV at the level of 1800~V where the avalanche was not large enough to give detectable light on the SiPM, and HV of about 2200~V where a clear avalanche light signal was observed as explained in Sec.~\ref{sec:AvalancheAnalysis}. Those two HV regimes correspond to a detector gain of about 15 and 50, respectively.


\subsection{SiPM calibration}

\begin{figure}[thp]
\centering
\includegraphics[width=0.75\textwidth]{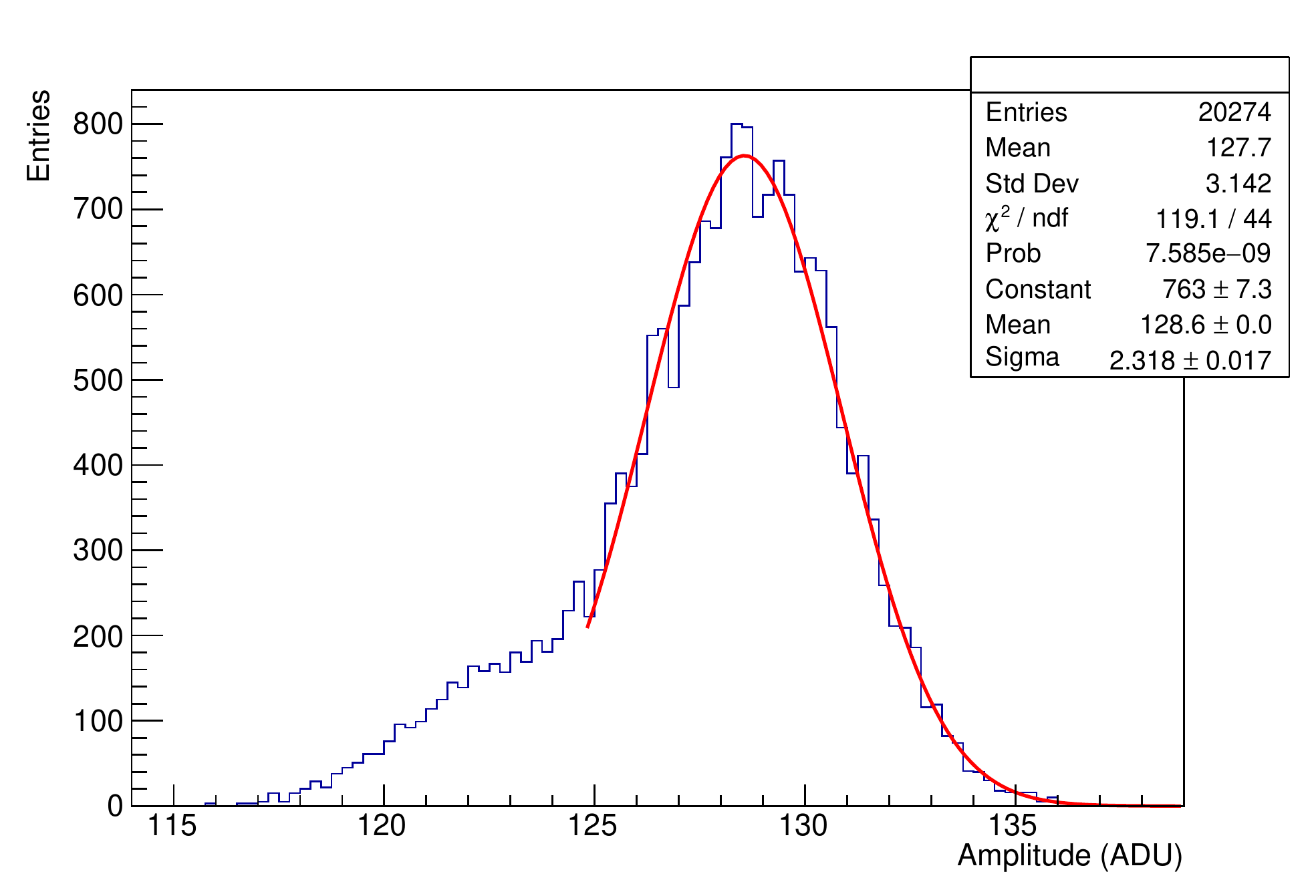} 
\caption{{\it Signal recorded by the same readout electronic chain used for the SiPM for a 10~mV pulse.}\label{fig:fig9}}
\end{figure}

\begin{figure}[thp]
\centering
\includegraphics[width=0.75\textwidth]{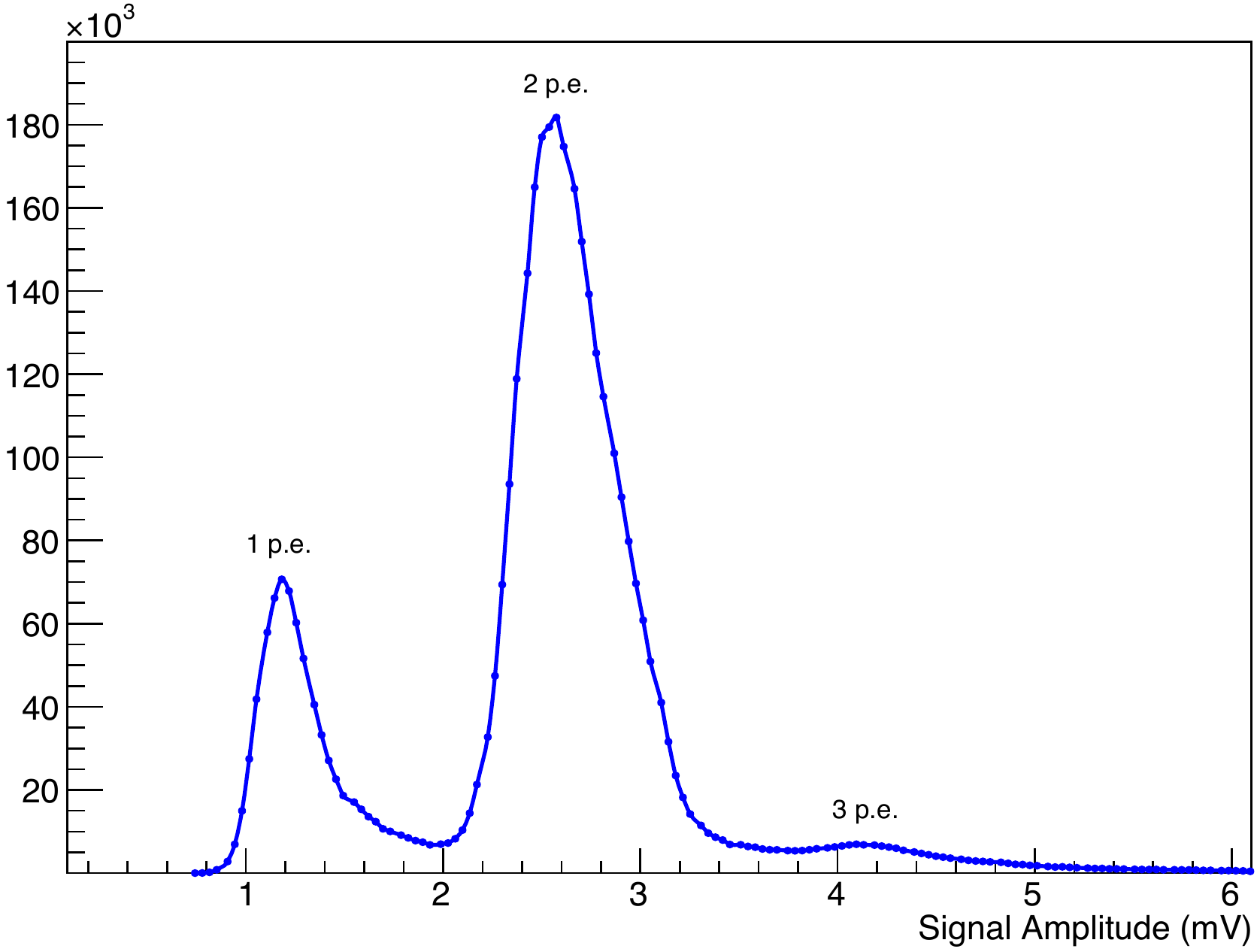} 
\caption{{\it Amplitude of SiPM dark noise signals. The 1~p.e. and 2~.p.e. peaks are clearly visible.}\label{fig:fig10}}
\end{figure}

The first point to be addressed was the SiPM calibration. Considering that the SiPM was operated in an unusual condition, at a large distance from the readout card, the noise prevented having the typical SiPM charge spectrum where the different number of photoelectrons are clearly seen.\\
Another reason why the typical SiPM charge spectrum was not observed is that the detector DAQ uses a CALI card~\cite{Armengaud:2017rzu} at a sampling rate of 2~MHz (i.e. bins of 500~ns) which records the SiPM signal of $\sim 200$~ns in a single time bin. The signal amplitude is therefore smeared by the coarse time sampling transforming the typical ``comb-like'' structure of the SiPM readout into a continuous distribution.

To calibrate the SiPM a two-step procedure was therefore applied. First a defined pulse of 10~mV, from a KEYSIGHT 33600A waveform generator, was used as input in the same readout electronics chain used for the SiPM. In order to account for the slow sampling rate of the DAQ with respect to the signal width, a signal similar to the real one was used, namely a pulse with a rise-time of 70~ns and a fall-time of 118.8~ns at a rate of 1~kHz. The obtained distribution shown in Fig.~\ref{fig:fig9} gives a conversion factor of 1~mV equal to $12.9$ Data Acquisition Units (ADU). Linearity was also checked and demonstrated within the input uncertainty between 10 and 30~mV.\\
The second step consisted of removing the SiPM from the detector and measuring its dark noise signal with the same readout electronics but directly connected to the readout circuit in order to remove the noise due to the cables. This was done using a Tektronix oscilloscope in order to retrieve the expected charge readout pattern. A random trigger was used and the maximum amplitude of the waveform in a window of about 700~ns before the trigger was recorded. The amplitude of the signal corresponding to a given number of photoelectrons is not affected by the window width, however the probability of having a given number of photoelectrons does depend on it. The larger the window the higher the probability of having a large signal corresponding to 2 photoelectrons or more.
The results is shown in Fig.~\ref{fig:fig10} where peaks corresponding to 1 and 2 photoelectrons are clearly visible: the amplitude corresponding to 1 p.e. corresponds to about 1.2~mV. The pedestal is not seen since it is below threshold.
A conversion can therefore be established between and DAQ units, such that 1 photoelectron corresponds to about 15.5 ADU.

\subsection{Drift time analysis}

\begin{figure}[thp]
\centering
\includegraphics[width=0.75\textwidth]{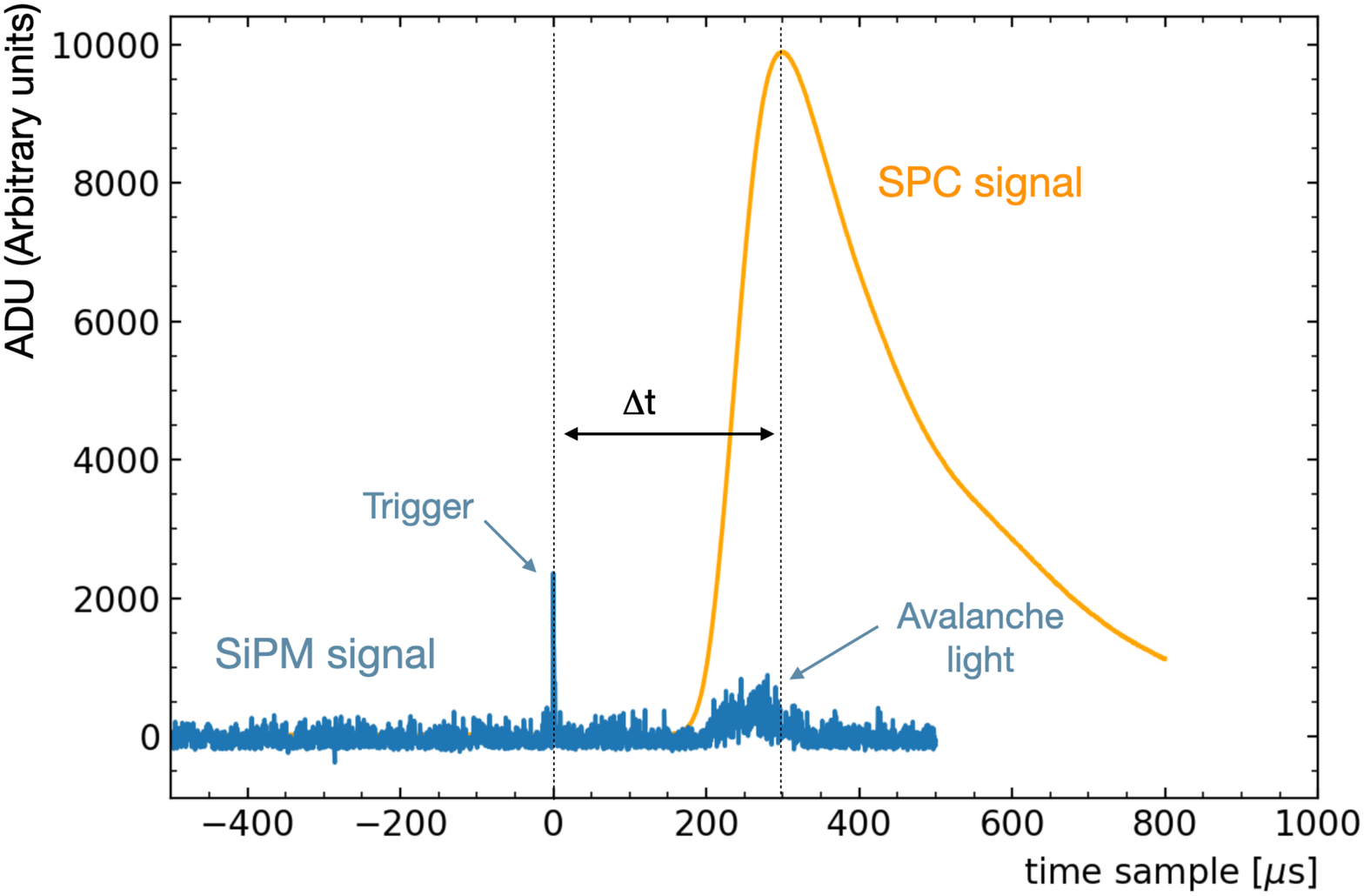} 
\caption{{\it Raw waveforms of SiPM (blue) and anodie signal (orange) for one event taken at 1.1~bar and 2200~V. The $\Delta t$ of about 300~$\mu$s is shown. For illustration purpose the SiPM signal is multiplied by a factor of 5.}\label{fig:fig5}}
\end{figure}

\begin{figure}[thp]
\centering
\includegraphics[width=0.7\textwidth]{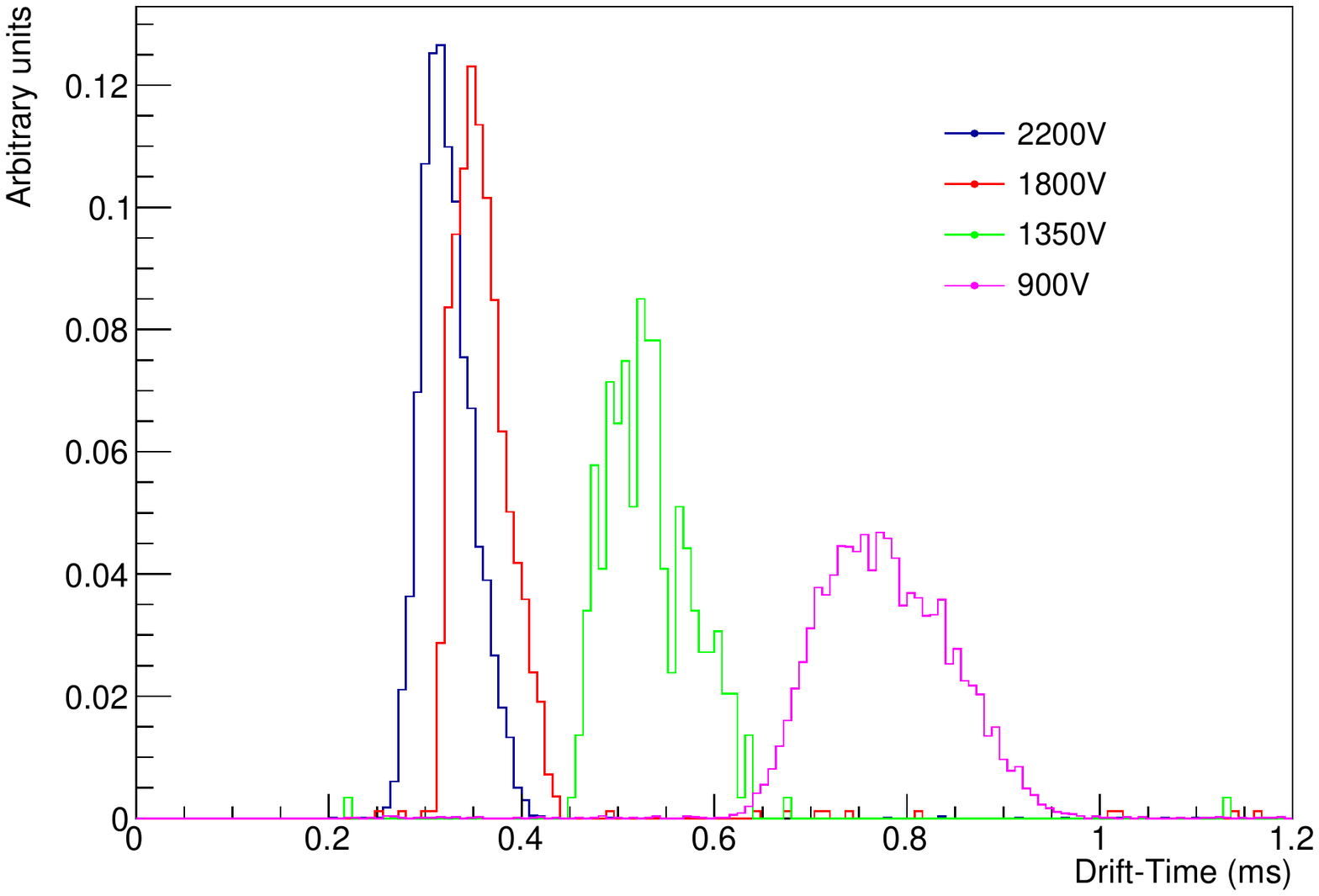} 
\caption{{\it Drift time obtained applying different HV on the central anode. The different histograms are normalized to one for a direct comparison independently on the number of triggered events of the run.}\label{fig:fig6}}
\end{figure}

\begin{figure}[thp]
\centering
\includegraphics[width=0.75\textwidth]{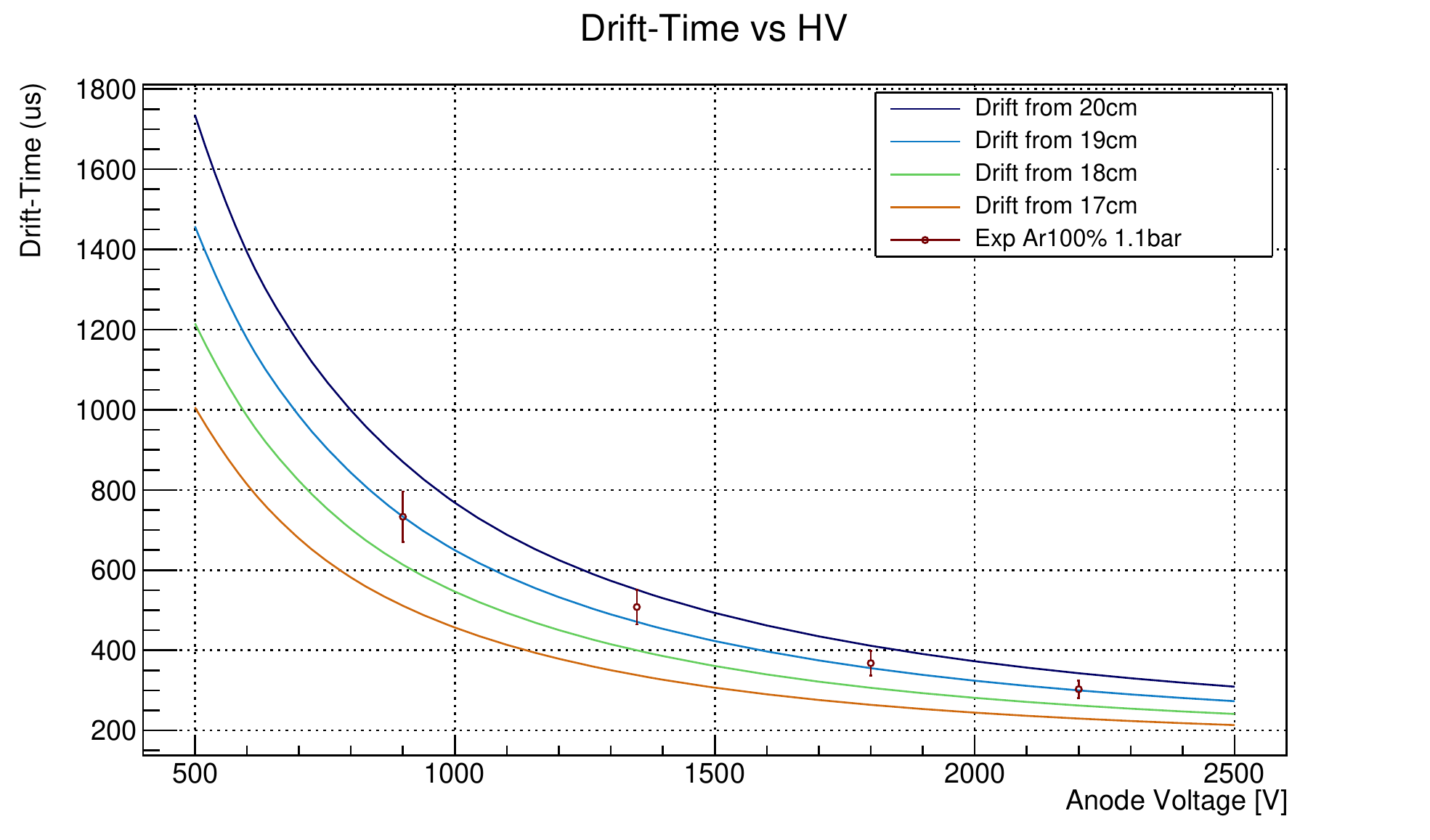} 
\caption{{\it Drift time from Garfield++ simulations for as a function of the anode HV for different starting radial distances from 17 to 20~cm (colored solid lines). The experimental data are also shown and the bar width corresponds to the RMS of the drift time distributions of Fig.~\ref{fig:fig6}.}\label{fig:fig7}}
\end{figure}

\begin{figure}[thp]
\centering
\includegraphics[width=0.7\textwidth]{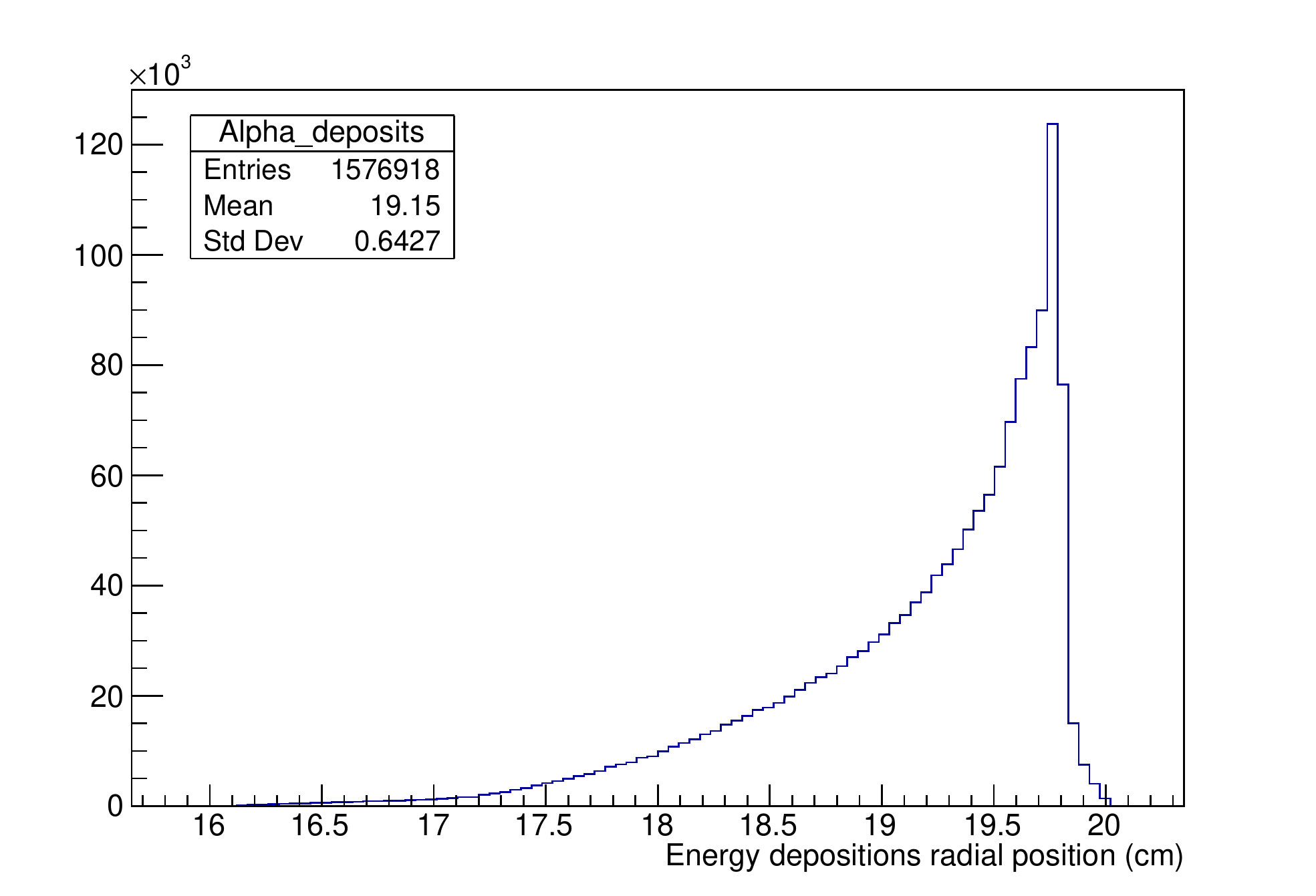} 
\caption{{\it Radial energy depositions from $\alpha$-particles simulated by Geant4 at 1.1~bar in argon.}\label{fig:fig8}}
\end{figure}

The first dataset was taken triggering on the SiPM with a threshold of 200 ADU which corresponds to about 13 p.e. At the time of measurements, the $^{210}$Po source had an activity of 0.4~Bq but only tracks passing in front of the SiPM produced detectable light; selecting this subsample of events reduced the trigger rate by a factor of 10 to 40 mHz.

Triggering on the SiPM selects a subsample of $\alpha$ tracks in a specific direction, however all these events should result in a signal on the central anode. This is indeed what we observed and one example of the waveforms for such events is shown in Fig.~\ref{fig:fig5}.

The time difference $\Delta t$ between the SiPM signal and the SPC signal was computed and is shown in Fig.~\ref{fig:fig6}. The width of the $\Delta t$ distribution depends on two factors: the spread due to the electron diffusion during the drift, and the $\alpha$ tracks emission angle. Considering that $\alpha$-particle tracks have a range of about 3~cm at 1.1~bar, the ionization electron distance from the anode spans between 17~cm and 20~cm.

A Garfield++~\cite{Veenhof:1998tt} simulation was carried out in order to benchmark the drift velocity of electrons. The drift time for different values of the anode HV was computed starting from radial positions spanning from 17~cm to 20~cm with respect to the detector center. The results are shown in Fig.~\ref{fig:fig7} along  with the measured data, indicating a mean distance of the $\alpha$-particle tracks of about 19~cm from the detector centre. This was confirmed by a Geant4~\cite{Agostinelli:2002hh} simulation of the $\alpha$ tracks: the radial position of the energy depositions exhibits a peak at 19.76~cm with a mean at 19.15~cm as shown in Fig.~\ref{fig:fig8}.


\subsection{Avalanche light analysis}
\label{sec:AvalancheAnalysis}

\begin{figure}[thp]
\centering
\includegraphics[width=0.7\textwidth]{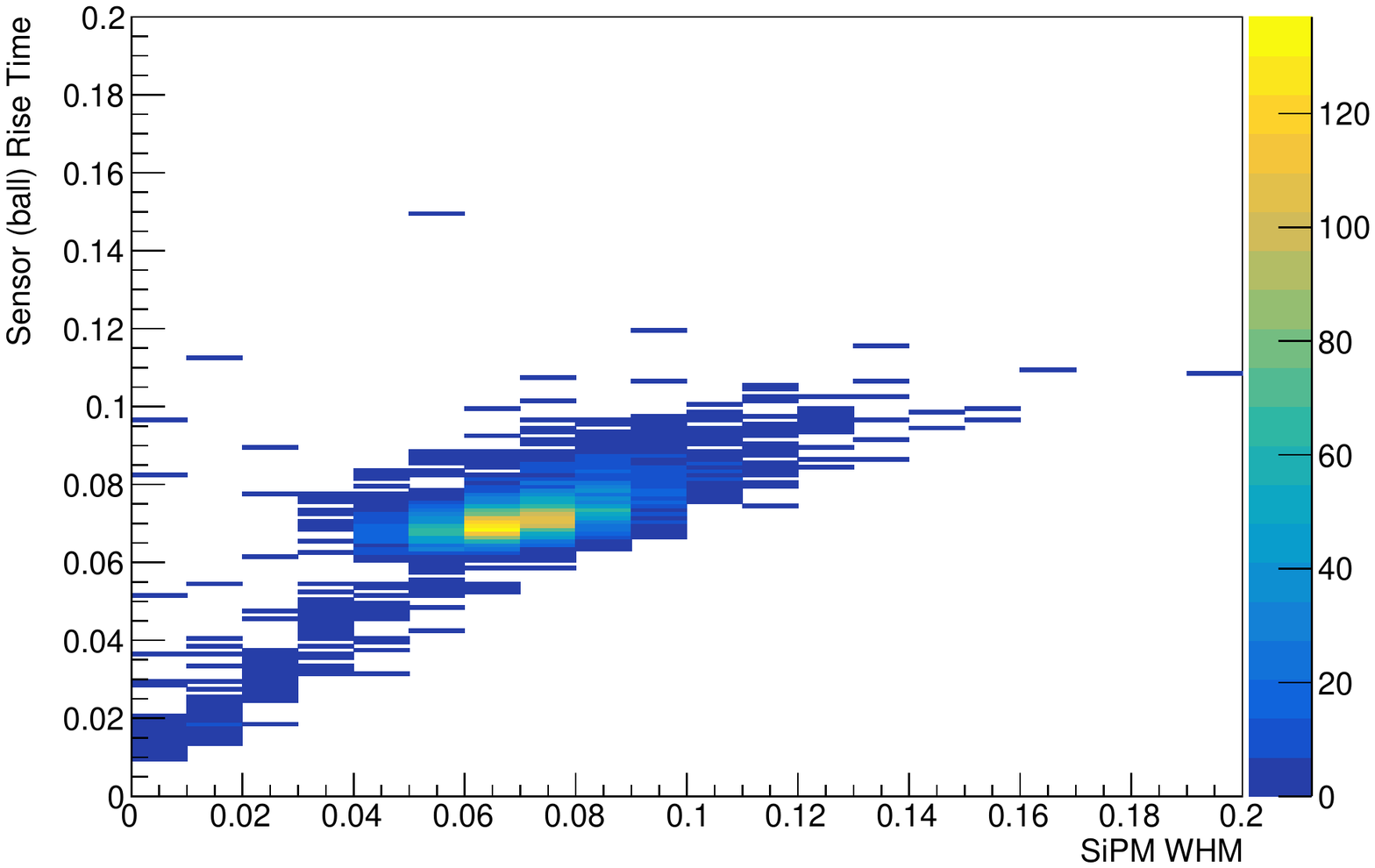} 
\caption{{\it Width of the avalanche light signal versus rise-time of the central anode signal.}\label{fig:fig11}}
\end{figure}

When the detector is operated at sufficiently high voltage, 2200~V in our case considering the argon gas pressure of 1.1~bar, the light signal emitted during the avalanche is large enough to be detected by the 6$\times$6 mm$^2$ SiPM located at the cathode surface. Such a light is emitted at the central anode and is observed also for $\alpha$ tracks which do not pass in front of the SiPM. It is therefore possible to trigger on the avalanche light to have a higher rate of events, at the level of 0.2~Hz expected from the source activity, and the trigger signal is in this case simultaneous with the central anode signal independently on the drift time.\\
Triggering on the avalanche light is of course not helpful for the radial track reconstruction, but such a signal could potentially be exploited to have an additional handle to reconstruct the topology of the event. The width of the waveform is indeed proportional to the time between the first and the last drifted electrons which in turn depends on the radial projection of the track. To validate such a feature we compared the width of the avalanche light signal to the rise-time of the central anode signal which is known to provide information on the radial position of the energy deposits as demonstrated by the NEWS-G collaboration and previous R2D2 works~\cite{Arnaud:2018bpc,Bouet:2020lbp}. The results shown in Fig.~\ref{fig:fig11} show a clear correlation between the two variables demonstrating that the avalanche light waveforms contains indeed information on the event topology. A dedicated analysis to exploit this feature will be the topic of future work.

\subsection{Secondary electrons signal}
\begin{figure}[thp]
\centering
\includegraphics[width=0.7\textwidth]{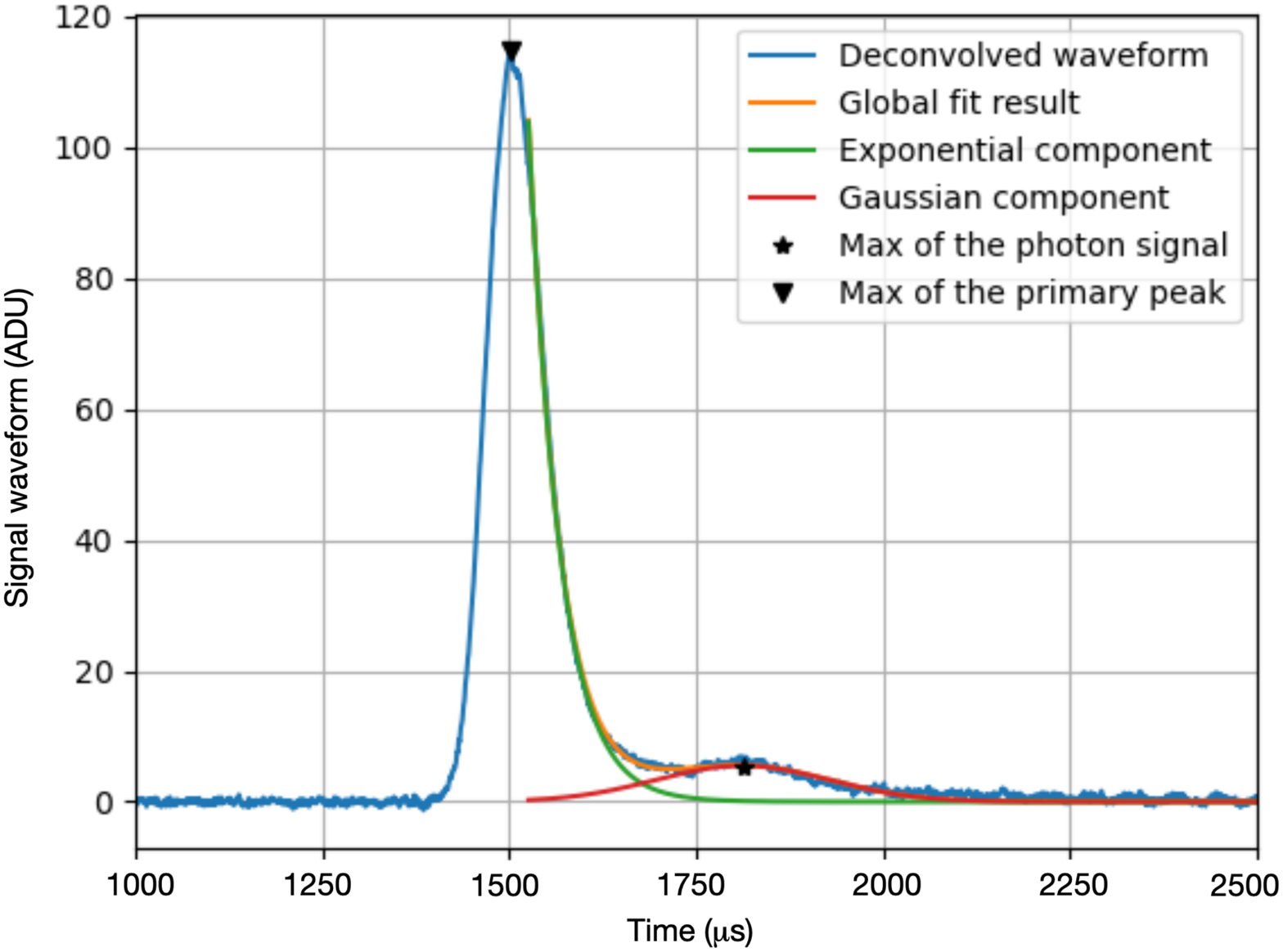} 
\caption{{\it Waveform, after signal processing (deconvolution), showing the bump due to the secondary electrons. The bump is modelled as a Gaussian function (red curve) over an exponential function (green curve) corresponding to the tail of the primary ionization peak.}\label{fig:fig12}}
\end{figure}

\begin{figure} [t]
\centering
\subfigure[\label{fig:fig13a}]{\includegraphics[height=5.
cm]{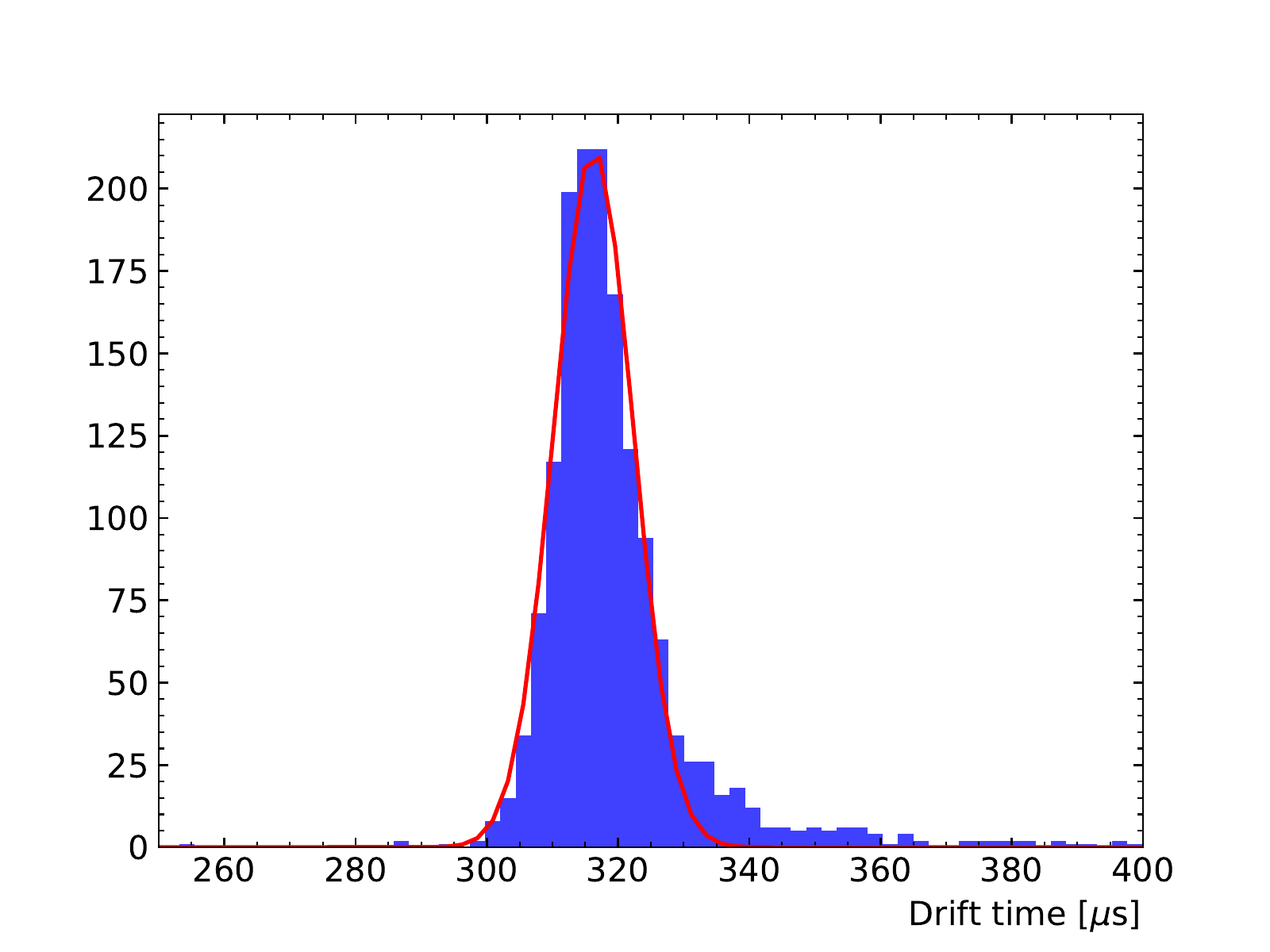}}
\subfigure[\label{fig:fig13b}]{\includegraphics[height=5.
cm]{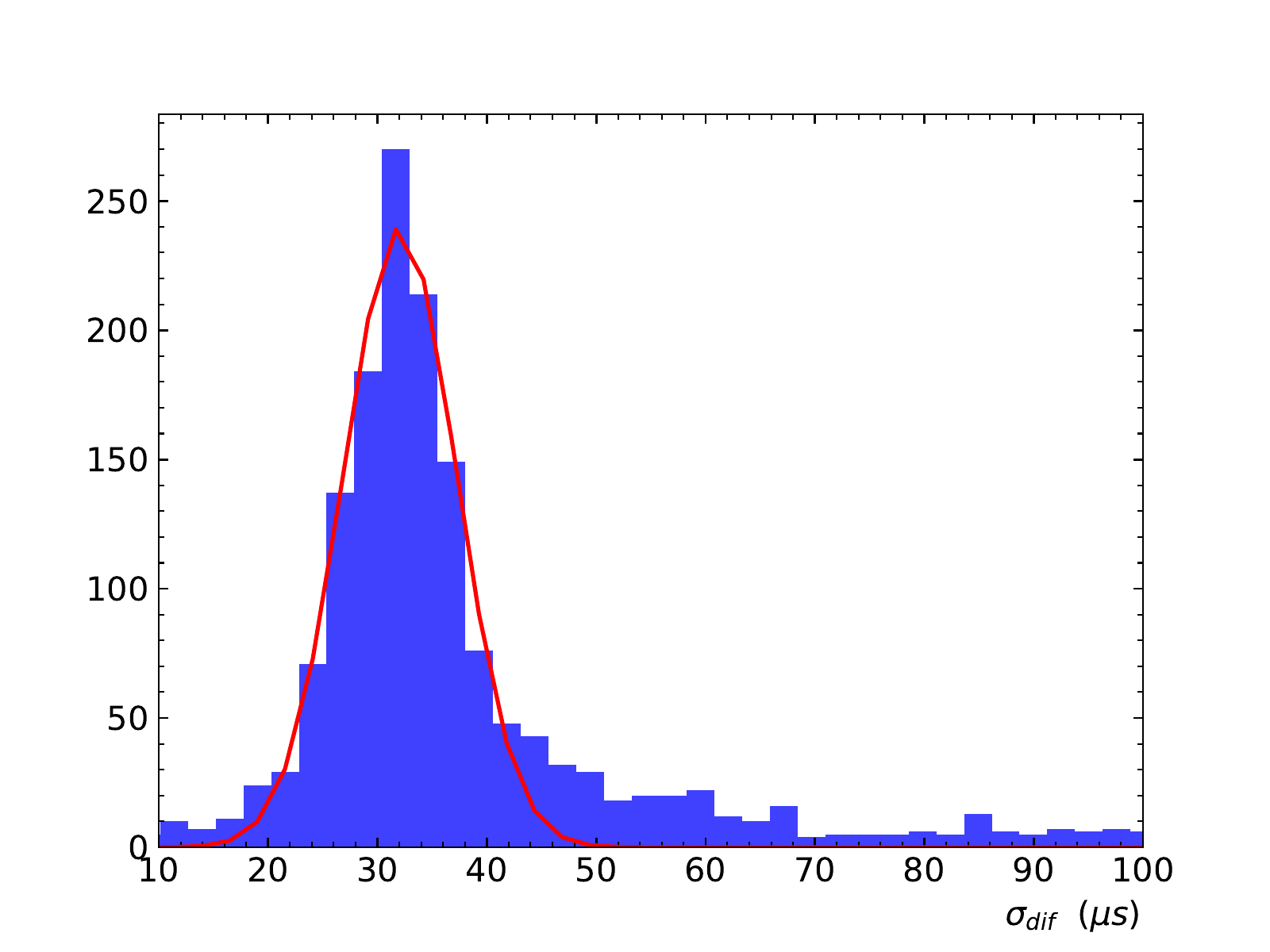}} \caption{{\it \subref{fig:fig13a} Time difference between the primary ionization signal and the bump due to secondary electron signal.The red is a Gaussian fit. \subref{fig:fig13b} Distribution of the width difference ($\sigma_{dif}$) between the peak of primaries and the induced photon peak, fitted by a Gaussian.}}
\label{fig:fig13}
\end{figure}

\begin{figure}[thp]
\centering
\includegraphics[width=0.7\textwidth]{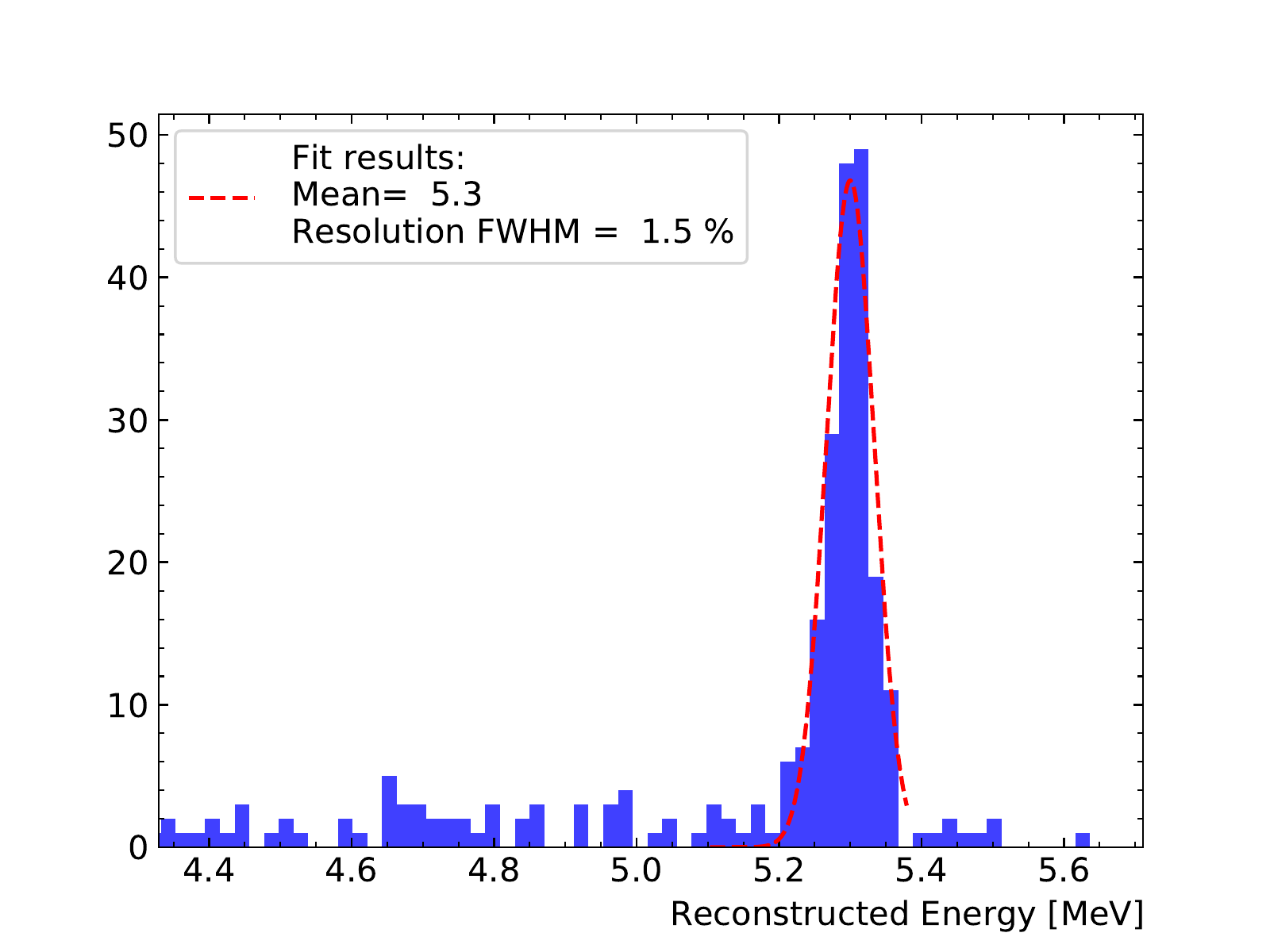} 
\caption{{\it Resolution obtained integrating the waveform over the full range.}
\label{fig:fig14}}
\end{figure}

The runs taken at 2200~V on the central anode, with a significant amount of light emitted in the avalanche process, allowed to study another phenomenon, namely the secondary electrons signal.\\
Photons produced during the avalanche reach the sphere surface and could eject electrons which would be drifted to the central anode producing a second signal. A digital devonvolution of the charge preamplifier output waveform allowed to recover the charge time distribution which was masked by the RC effect of the charge integrator (see~\cite{Bouet:2020lbp} for the method). The resulting deconvolved waveform exhibits a bump on the tail of the primary ionization signal, as seen in Fig.~\ref{fig:fig12}. . 
For further analysis, this photon-induced contribution was modeled with a Gaussian distribution superimposed on an exponential component describing the tail of the primary ionization peak. 
The time difference between the secondary electron signal and the primary ionization one was computed as shown in Fig.~\ref{fig:fig13a}. The mean value at about 316~$\mu$s corresponds indeed to the drift time expected from electrons starting at the sphere surface (i.e. distance of 20~cm from the anode) for a central anode at 2200~V as shown in Fig.~\ref{fig:fig7}.\\
It is possible to extract the value of the signal time spread due to diffusion ($\sigma_{dif}$). Indeed if the electrons were all produced at the same time the Gaussian width would be completely due to the electron diffusion. This is not the case and the time spread of the production is given at first approximation by the time spread of the primary ionization signal. The diffusion can therefore be evaluated as: $\sigma_{dif} = \sqrt{\sigma_{photon}^2 - \sigma_{ion}^2}$ where $\sigma_{photon}$ and $\sigma_{ion}$ are the widths of the Gaussian fits of the secondary electron signal and of the primary ionization signal respectively. The mean value of the diffusion over the full 20~cm drift is 32~$\mu$s (see Fig.~\ref{fig:fig13b}).\\
It is noted that in case of high gain operation, secondary electron emissions could become challenging since it may bring the detector into a Geiger mode. The addition of a quencher to the gas allows to mitigate this effect, however, in a future Xe-filled SPC, the presence of a quencher could result in an additional degree of complication concerning xenon purity since the use of a purifier, such as a hot getter, could modify the quencher fraction over time.
For this reason, considering the foreseen low gain operation, the use of pure xenon is currently preferred. 
The impact on the energy resolution of the secondary electron signal was therefore an important point to be addressed.
The waveform integral was studied, for fully contained tracks,
and a resolution of 1.5\% FWHM was found at 5.3~MeV as shown in Fig.~\ref{fig:fig14}. If the secondary electron signal is subtracted (modelled with a Gaussian function) the resolution is degraded to 2.2\%.\\
Such a result is of great interest for the development of the future detector for $\beta\beta0\nu$ decay search since it suggests that pure xenon may be used also in proportional mode and the use of a quencher is not mandatory to achieve the desired energy resolution (at least within our modest gain operating regime, and the high energy deposits expected with $\beta\beta0\nu$ search).\\
This is true for what concerns the use of quencher in order to avoid secondary electrons stripped from the sphere by scintillation light. 
This result illustrates the possibility to operate the detector without quencher, but the quencher presence has also an impact on the resolution, affecting the number of collected electron via Penning effect~\cite{Sahin:2010ssz}.
Indeed, previous results shown a better resolution with 98\%Ar:2\%CH$_4$~\cite{Bouet:2020lbp} with respect to the resolution presented here, obtained in pure argon.

\section{Conclusions}

For the first time a scintillation light signal is used as trigger in a SPC. 
The presented results, although still partial, constitute an encouraging first step towards the use of an SPC filled with pure noble gas. Studies on the waveform of the light signal have been carried out and the possibility of crossing information provided by the anode signal was investigated. 
Observations showed an excellent agreement between the ionization electron drift time and the expectation for the GARFIELD++ simulations.

An additional important outcome of this study is that, in a suitable regime of gain, the use of a quencher to avoid secondary electrons is not mandatory, at least at the tested pressure. The presence of secondary electrons emitted from the cathode by photons produced in the avalanche, does not spoil the detector energy resolution.

Further work is required to validate these findings for a xenon high pressure detector aiming at the detection of $\beta\beta0\nu$. 

\section*{Acknowledgments}
 The authors would like to thank the IdEx Bordeaux 2019 Emergence program for the OWEN grant for the ``Development of a custom made electronics for a single channel time projection chamber detector aiming at the discovery of neutrinoless double beta decays, and for possible applications in industry''. In addition we thank the CNRS International Emergency Action (IEA) for the ``E-ACHINOS'' grant supporting the collaboration between CENBG and University of Birmingham. We thank the CENBG technical staff. We thank M.~Chapellier for providing the Po source and for useful discussions.
This project has received funding from the European Union's Horizon 2020 research and innovation programme under the Marie Sk\l{}odowska-Curie grant agreement no 841261 (DarkSphere). K.~Nikolopoulos acknowledges support by the European Research Council (ERC).

\bibliography{references}

\end{document}